\documentclass[fleqn,usenatbib]{mnras}

\usepackage{newtxtext,newtxmath}
\usepackage[T1]{fontenc}

\DeclareRobustCommand{\VAN}[3]{#2}
\let\VANthebibliography\thebibliography
\def\thebibliography{\DeclareRobustCommand{\VAN}[3]{##3}\VANthebibliography}

\usepackage{graphicx}	% Including figure files
\usepackage{amsmath}	% Advanced maths commands
\title[$z\gtrsim10$ galaxies in the HUDF]{The abundance of $\mathbf {z \gtrsim 10}$ galaxy candidates in the HUDF using deep JWST NIRCam medium-band imaging}

\author[C.\,T. Donnan et al.]{C.\,T. Donnan$^{1}$\thanks{E-mail: callum.donnan@ed.ac.uk}
D.\,J. McLeod$^{1}$,
R.\,J. McLure$^{1}$,
J.\,S. Dunlop$^{1}$,
A.\,C. Carnall$^{1}$,
F. Cullen$^{1}$, 
D.\,Magee$^{2}$
\\
$^{1}$Institute for Astronomy, University of Edinburgh, Royal Observatory, Edinburgh, EH9 3HJ. UK\\
$^{2}$ Department of Astronomy and Astrophysics, UCO/Lick Observatory, University of California, Santa Cruz, CA 95064, USA\\
}

\date{Accepted XXX. Received YYY; in original form ZZZ}

\pubyear{2022}

\begin{document}
\label{firstpage}
\pagerange{\pageref{firstpage}--\pageref{lastpage}}
\maketitle

\begin{abstract}
We utilise \textit{JWST} NIRCam medium-band imaging to search for extreme redshift ($z\geq9.5$) galaxy candidates in the Hubble Ultra Deep Field (HUDF) and the additional pointing within the GOODS-South field provided by the second NIRCam module. Our search reveals 6 robust candidates, 3 of which have recently been spectroscopically confirmed.  One of these 3 is the previously controversial $z\simeq 12$ galaxy candidate UDF-22980 which is now detected in five \textit{JWST} NIRCam medium-band filters (F182M, F210M, F430M, F460M and F480M), efficiently excluding alternative low-redshift solutions and allowing us to now report a secure photometric redshift of $z_{\rm{phot}}=11.6\pm0.2$. We also detect 2 galaxies at $z\geq12.5$ including a newly-detected candidate in the imaging provided by the second NIRCam module (south-west of the HUDF) at $z_{\rm{phot}}=12.6\pm0.6$. We determine the physical properties of the 6 galaxies by fitting the 14-band photometry with \textsc{Bagpipes}. We find stellar masses of $\log(M_{\star}/{\rm M_{\odot}}) \simeq 7.5 - 8.7$ and star-formation rates of $\log(\rm{SFR}/M_{\odot}^{-1} \rm{yr}^{-1}) \simeq 0.3 - 5.0$. Despite the relatively small cosmological volume covered by the HUDF itself and the second NIRCam module imaging, we find that the existence of these galaxies is fully consistent with the latest measurements of both the UV luminosity function and cosmic star-formation rate density at $z\simeq11$, supporting a gradual steady decline in the cosmic star-formation rate density out to at least $z\simeq15$. 
\end{abstract}


%%%%%%%%%%%%%%%%%%%%%%%%%%%%%%%%%%%%%%%%%%%%%%%%%%

%%%%%%%%%%%%%%%%% BODY OF PAPER %%%%%%%%%%%%%%%%%%
\section{Introduction}
The study of galaxies at the highest redshifts is crucial to unveiling the earliest stages of galaxy formation and evolution. Observations of high-redshift galaxies are vital for improving our understanding of the formation of the first stars and black holes and quantifying the role of galaxies in cosmic reionization (see \citealt{dunlop2013} and
\citealt{stark2016} for reviews). 

Until recently, the observational frontier was defined by galaxies that could be selected and studied using a combination of deep near-infrared imaging with \textit{HST} and shallower mid-infrared imaging with \textit{Spitzer}. This combination of facilities led to significant improvements in our understanding of galaxy evolution out to $z\simeq 9$ \citep[e.g.][]{ellis2013,mclure2013,oesch2014,oesch2018,finkelstein2015,mcleod2015,mcleod2016,bouwens2021,bouwens2022}. Due to its long-wavelength limit of 
$1.7\, \mu$m, \textit{HST} imaging can only identify galaxies out to $z\simeq 12$, and even then only with single-band detections in the F160W filter. Prior to the launch of the {\it James Webb Space Telescope} ({\it JWST}), the most distant  spectroscopically-confirmed galaxy had a redshift of $z=10.957 \pm 0.001$: GN-z11  \citep{oesch2016,jiang2021}, a source that was originally selected using {\it HST} F160W imaging combined with non-detections at shorter wavelengths.

The landscape of high-redshift galaxy studies has now been dramatically changed with the advent of {\it JWST}, which has enabled a new era of extragalactic astronomy. The extended wavelength coverage of NIRCam imaging to $\lambda \simeq 5\,\mu$m has facilitated the robust detection of galaxies at $z\geq10$ for the first time \citep[e.g.][]{finkelstein2022,finkelstein2022c,adams2022b,castellano2022,naidu2022} including galaxy candidates at $z\simeq16$ \citep[][]{donnan2022,harikane2022}. Early {\it JWST}-based studies of the evolving galaxy UV luminosity function imply a smooth and steady decline in the cosmic star-formation rate density ($\rho_{\rm{SFRD}}$) out to $z\simeq15$ \citep{donnan2022,harikane2022}, as anticipated from some previous \textit{HST}-based studies \citep[e.g.][]{mcleod2016}.  

Although GN-z11 was the highest redshift, spectroscopically-confirmed galaxy prior to {\it JWST}, another even higher-redshift galaxy candidate was identified in the \textit{HST} UDF12 imaging of the HUDF \citep[][]{ellis2013,mclure2013} with a best-fitting redshift of $z_{\rm{phot}}=11.9^{+0.3}_{-0.5}$. This galaxy (UDF12-3954-6284) is the highest-redshift candidate ever identified from \textit{HST} imaging. In fact it was originally noted as a potential $z\simeq 10$ galaxy (UDFj-39546284) from the UDF09 imaging of the HUDF \citep[][]{bouwens2011}, based on a F160W detection and non-detections in F125W, F105W and the HUDF ACS optical imaging. However, following the UDF12 {\it HST} imaging campaign, the galaxy was also found to be undetected in the F140W filter which, given the substantial overlap between the F140W and F160W filters, favoured a significantly higher redshift of $z\simeq 11.9$ \citep[][]{ellis2013}. 

\citet{robertson2022} and \citet{curtislake2022} have recently reported {\it JWST} NIRCam and NIRSpec observations of 4 galaxies within the GOODS-S field from the JADES survey, including spectroscopic confirmation of UDFj-39546284 at $z_{\rm{spec}}=11.58\pm0.05$. They also report a galaxy at $z_{\rm{spec}}=13.20^{+0.04}_{-0.07}$ which now overtakes Gn-z11 as the highest redshift spectroscopically-confirmed galaxy. In addition, they derive stellar masses, star-formation rates and mean stellar ages for these 4 objects, finding results consistent with expectations at these redshifts.

\citet[][]{bouwens2022} used \textit{JWST} NIRCam medium band imaging to measure the evolution of the UV luminosity function in the redshift range $z=8$-$15$ using high-redshift galaxy candidates selected from within the HUDF area alone. Consistent with other studies \citep[e.g.][]{donnan2022,harikane2022} their results do not support the previously suggested rapid decline in the cosmic star-formation rate density ($\rho_{\rm{SFR}}$) beyond $z\sim8$ \citep{oesch2018}. However, their measurements of $\rho_{\rm{SFR}}$ at $z \gtrsim11$ are in fact significantly higher than what was derived by \citet{donnan2022}. They also measure an absolute UV magnitude, $M_{\rm{UV}}$, that is $\simeq 1$ magnitude brighter than reported by \citet{robertson2022} for the spectroscopically confirmed $z_{\rm{spec}}=11.58\pm0.05$ source. 

Motivated by the continued uncertainty over the nature of the UV luminosity function and cosmic star-formation rate density at $z\gtrsim10$, we analyse a combination of the ultra-deep \textit{HST} ACS+WFC3/IR data available in GOODS-S/UDF with the full area available from \textit{JWST} NIRCam medium-band imaging at $\lambda \simeq 2\,\mu$m and $\lambda~\simeq~4.5\,\mu$m to search for galaxy candidates at $z\geq9.5$. Crucially, the new NIRCam imaging (ID 1963; PI Williams) is the first publicly available data in the HUDF red-ward of $\lambda \simeq1.6\, \mu$m with sufficient sensitivity to robustly detect candidates at $z\geq9.5$. 
The aim of this study is to exploit the new \textit{JWST} imaging data to derive secure photometric redshifts for $z\geq9.5$ galaxy candidates and provide further constraints on the galaxy UV luminosity function and hence luminosity density at $z\geq9.5$. We also employ Spectral Energy Distribution (SED) fitting to derive the basic properties of the stellar populations in these early galaxies.

The paper is structured as follows. In Section \ref{sec:data} we describe the imaging data and source catalogue creation. In Section \ref{sec:results} we explain the sample selection and the measurement of the physical properties of the galaxies. In Section \ref{sec:LF} we present our derived luminosity function and the resulting inferred high-redshift dependence of the cosmic star-formation rate density. In Section \ref{sec:discussion} we discuss our results in the context of other recent observational studies and the predictions of various theoretical/numerial models of galaxy formation. Finally, in Section \ref{sec:conclusions} we summarise our conclusions. Throughout we use magnitudes in the AB system \citep{oke1974,oke1983}, and assume a standard cosmological model with $H_0=70$ km s$^{-1}$ Mpc$^{-1}$, $\Omega_m=0.3$ and $\Omega_{\Lambda}=0.7$.

\section{Data}
\label{sec:data}

\subsection{Imaging}
We utilise deep imaging from \textit{HST} and \textit{JWST} in the HUDF. From \textit{HST} we use data covering the wavelength range $\lambda \simeq 0.4 - 1.6\,\mu m$ from ACS F435W, F606W, F775W, F814W, F850LP and WFC3 F105W, F125W, F140W and F160W imaging. We utilise new \textit{JWST} imaging from the HUDF medium-band survey (ID 1963; PI Williams) which contains deep NIRCam imaging through the F182M, F210M, F430M, F460M and F480M filters \citep{williams2023}. The $5\sigma$ global depths for F182M and F210M are $m_{\rm{AB}}\simeq 29$ and for F430M, F460M and F480M they are $m_{\rm{AB}}\simeq 28.5$. The data were reduced using the Primer Enhanced NIRCam Image Processing Library (PENCIL; Magee et al., in prep) software, a customised version of the \textit{JWST} pipeline v1.8.0. The CRDS context used was ``jwst\_1011.pmap'', which includes the most recent NIRCam zero-point corrections and in-flight calibrations. An extended description is available in \citet{donnan2022}.

\subsection{Catalogue creation and photometry}
We perform a search for galaxies using \texttt{SExtractor} \citep[][]{bertin1996} in dual-image mode. 
We use the F182M band image as the detection image as it is the deepest imaging sampling flux long-ward of the Lyman break. For the photometry in the {\it JWST} NIRCam imaging, we use a similar technique to \citet{donnan2022}, with 0.24\,arcsec-diameter apertures in the SW filters (F182M, F210M) and 0.3\,arcsec-diameter apertures in the LW filters (F430M, F460M, F480M). We also use 0.3\,arcsec-diameter apertures in the HST ACS and WFC3 filters. All of the photometry was then corrected to 76\% of total (as this is approximately the flux enclosed by the apertures utilised for F182M) using a point-source correction based on the curve-of-growth of each filter's PSF. 
We adopt the $1\sigma$ local depth as the error on the fluxes measured through each filter. This was determined by $1.483 \times \rm{MAD}$ where $\rm{MAD}$ is the median absolute deviation of the flux in the closest 200 empty apertures to the source.

\begin{figure*}
	\includegraphics[width=\textwidth]{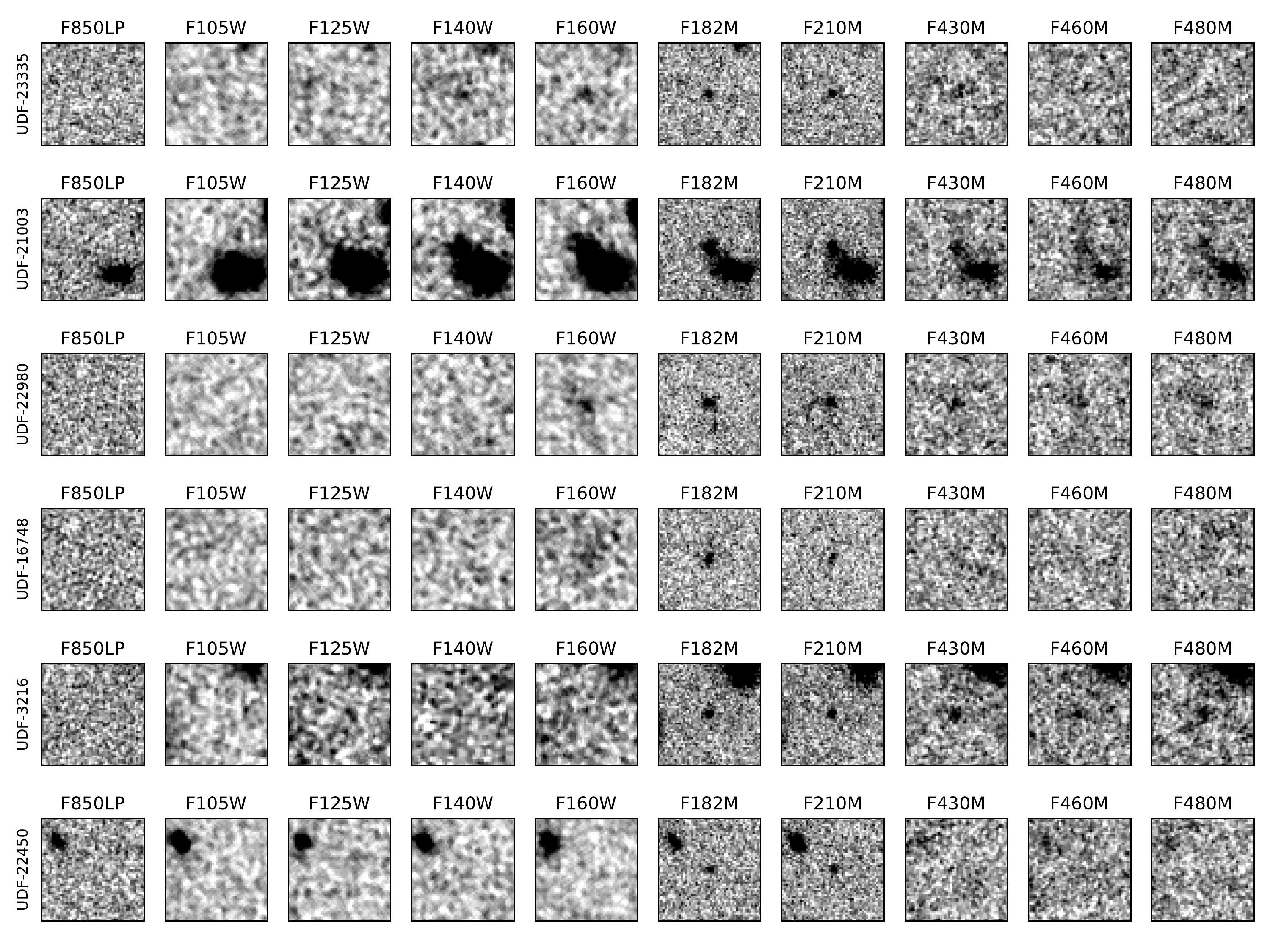}
    \caption{Postage-stamp greyscale images of the six $z\geq9.5$ galaxy candidates from our sample in the \textit{HST} and \textit{JWST} imaging. The imaging is ordered by increasing wavelength from left to right. Each postage-stamp image is 1.5 $\times$ 1.5\,arcsec in size, with North to the top and East to the left.}
    \label{fig:stamps}
\end{figure*}

\section{Galaxy sample}
\label{sec:results} 
\subsection{Sample selection}
We use the photometric redshift (photo-z) code \texttt{EAZY} \citep{brammer2008} for our primary redshift determination, fitting with the \textsc{Pegase} \citep{pegase1999} set of templates which include nebular emission lines. We determine the primary solution by allowing templates to vary over the redshift range $0<z<25$ and a secondary low-redshift solution by restricting the redshift range to $0<z<6$. We select our sample by requiring a best-fitting photometric redshift at $z\geq9.5$ with a threshold value of $\Delta \chi^2 \geq 4$ between the primary and secondary solution. Due to the position of the Lyman break, we require the flux to be $<2 \sigma$ in all the ACS filters. We also require the flux to be $>5 \sigma$ in either F160W or F182M with the next reddest filter at $3\sigma$ to ensure a robust detection long-ward of the Lyman break. We visually inspect the final sample to remove artefacts and diffraction spikes. To verify the robustness of the photometric redshifts, we also fitted the photometry using two further SED fitting codes: \texttt{LePhare} \citep[][]{arnouts1999,ilbert2006} with templates from \citet{bruzual2003} and with dust attenuation spanning the range $A_{V}=0.0-6.0$, and also the code described in \citet{mclure2011}. We require all candidates in the final sample to have significantly preferred high-redshift solutions from all three codes.

\subsection{Final sample}
\label{sec:sample}

\begin{table*}
	\centering
	\caption{The observed photometry for the galaxies in the {\it JWST}-selected sample in AB magnitudes. The first column lists the ID of the object. The following columns show the photometry in each of the relevant ACS, WFC3 and NIRCam filters. In the case of a non-detection at the 2$\sigma$ level the photometry is shown as an upper limit.}
	\def\arraystretch{1.25}% 
	\begin{tabular}{c | c | c | c | c | c | c | c | c | c | c}
		\hline
		ID & F850LP & F105W & F125W & F140W & F160W & F182M & F210M & F430M & F460M & F480M \\
		\hline
        23335 & >30.40 & >31.20 & 30.49$^{+0.45}_{-0.32}$ & 29.71$^{+0.18}_{-0.15}$ & 29.41$^{+0.14}_{-0.12}$ & 28.94$^{+0.14}_{-0.13}$ & 29.10$^{+0.21}_{-0.18}$ & 28.85$^{+0.41}_{-0.3}$ & >28.95 & >29.25 \\
		21003 & >30.41 & 30.68$^{+0.36}_{-0.27}$ & 29.27$^{+0.12}_{-0.1}$ & 28.09$^{+0.06}_{-0.05}$ & 27.77$^{+0.06}_{-0.05}$ & 27.84$^{+0.06}_{-0.05}$ & 27.94$^{+0.06}_{-0.05}$ & 28.00$^{+0.2}_{-0.17}$ & 27.89$^{+0.21}_{-0.17}$ & 28.26$^{+0.22}_{-0.18}$ \\
		22980 & >30.48 & >31.37 & >31.00 & >30.93 & 29.04$^{+0.1}_{-0.09}$ & 28.33$^{+0.09}_{-0.08}$ & 28.32$^{+0.08}_{-0.08}$ & 28.37$^{+0.2}_{-0.17}$ & 28.02$^{+0.19}_{-0.17}$ & 28.21$^{+0.23}_{-0.19}$ \\
		16748 & >30.16 & >30.93 & >30.50 & >30.54 & 29.68$^{+0.29}_{-0.23}$ & 28.93$^{+0.18}_{-0.15}$ & 29.10$^{+0.4}_{-0.29}$ & 29.15$^{+0.62}_{-0.39}$ & >29.24 & 29.24$^{+0.67}_{-0.41}$ \\
		3216 & >28.87 & >29.69 & >29.49 & >28.03 & >29.22 & 28.69$^{+0.1}_{-0.09}$ & 28.75$^{+0.13}_{-0.11}$ & 28.33$^{+0.23}_{-0.19}$ & 28.30$^{+0.25}_{-0.21}$ & 28.39$^{+0.15}_{-0.13}$ \\
		22450 & >30.28 & >31.39 & >30.69 & >30.99 & >31.01 & 29.29$^{+0.19}_{-0.16}$ & 28.99$^{+0.22}_{-0.18}$ & >29.16 & >29.20 & >29.26 \\
        \hline

	\end{tabular}
	\label{tab:jwst_phot}
\end{table*}

The final sample contains 6 galaxies at $z\geq9.5$. The measured photometry, corrected to total, is listed in Table \ref{tab:jwst_phot} and the photometric redshifts of the galaxy candidates are listed in Table \ref{tab:sample}. Out of these 6 sources, one is UDFj-39546284 (UDF-22980 in this study) which is the previously-detected source at $z\simeq12$ \citep[e.g.][]{bouwens2011,ellis2013} with recent spectroscopic confirmation at $z_{\rm{spec}}=11.58$ \citep{curtislake2022}. Our best-fitting photo-z for this galaxy from \texttt{EAZY} is in excellent agreement with this, yielding $z = 11.65^{+0.17}_{-0.21}$, with a $\Delta \chi^2=40$ between the primary solution and the low-redshift secondary solution ($z=2.9^{+0.1}_{-0.1}$), indicating a clear preference for the primary (higher) redshift. This is consistent with the photo-z from \citep{ellis2013,mclure2013} which at the time was based on only a F160W detection (combined with non-detections through other \textit{ HST} and \textit{Spitzer} filters). 

\begin{figure*}
	\includegraphics[width=\textwidth]{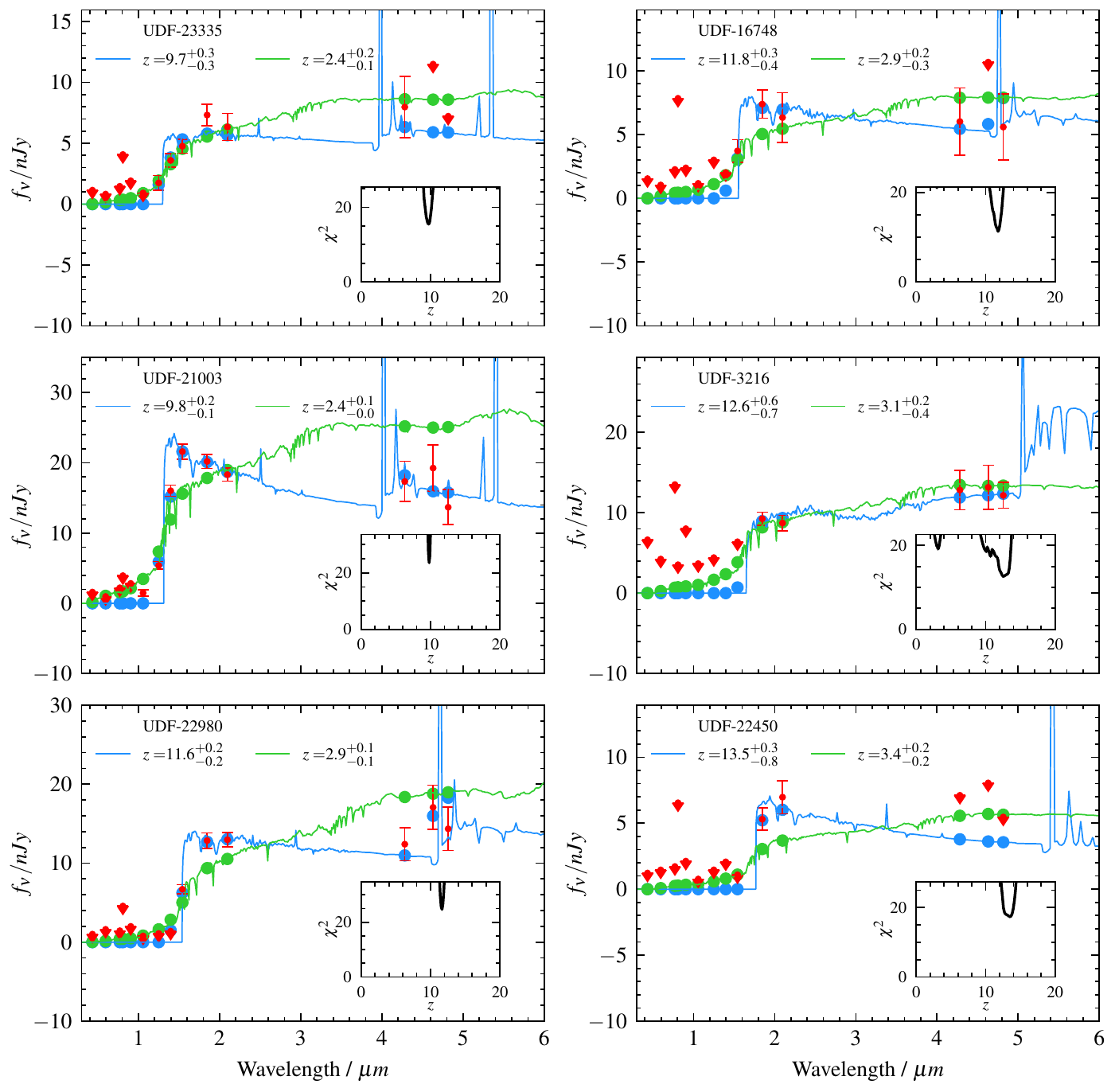}
    \caption{Alternative Spectral Energy Distribution (SED) fits for the six $z\geq9.5$ galaxies in the sample. The blue line shows the best-fitting (preferred) high-redshift solution, the green line shows the best-fitting (alternative) low-redshift solution, and the red points show the measured photometry (at 76$\%$ of total flux). The red arrows show upper limits in the case of a non-detection at the $2\sigma$ level. The solid blue and green circles represent the model photometry of the best-fitting high and low-redshift templates respectively. The value of $\chi^2$ as a function of redshift is shown for each source in the inset panel.}
    \label{fig:eazy_sed}
\end{figure*}

UDF-21003, with a photo-z of $z=9.8^{+0.2}_{-0.1}$, is a newly-detected source which is in fact the most robustly detected candidate in the sample. It was previously undetected likely due to the close positioning of a low-redshift nearby source. UDF-3216 is newly-detected candidate at $z=12.6\pm0.7$ which is located in the second/parallel NIRCam module. Multi-waveband postage-stamp images of the 6 galaxies in the sample are shown in Fig.\,1, while the photometric data for the sample along with the two alternative (high-redshift and low-redshift) SED fits for each galaxy are shown in Fig.\,\ref{fig:eazy_sed}. 

\begin{table*}
	\centering
	\caption{The best-fitting photometric redshifts and derived physical properties of the six $z\geq9.5$ galaxy candidates. The first column shows the ID of the object followed by the best-fitting photometric redshift from \texttt{EAZY}. The following columns show the UV slope ($\beta$), stellar mass, star-formation rate and mean stellar age from \textsc{Bagpipes}, the absolute UV magnitude and the coordinates of each object. The penultimate column notes if the source was also reported by \citet{bouwens2022} and the final column shows, where available, the spectroscopic redshift from \citet{curtislake2022}. }

	\label{tab:sample}
	\def\arraystretch{1.35}
	\begin{tabular}{lcccccccccc} % four columns, alignment for each
		\hline
		ID & $z_{\rm phot}$ & $\beta$  & $\log(M_{\star}/M_{\odot})$ & $\log(\rm{SFR}/{\rm M_{\odot}}/\rm{yr})$ & $t_{\star}$/Myr & $M_{\rm UV}$ & RA & Dec & B22 & $z_{\rm{spec}}$\\
		\hline
23335 & 9.68$^{+0.28}_{-0.28}$ & $-0.8\pm0.6$ & $7.5^{+0.4}_{-0.2}$ & $0.3^{+0.5}_{-0.1}$ & $7^{+38}_{-5}$ & $-18.13$ & 03:32:38.12 & $-$27:46:24.56 & Yes & 10.38  \\
21003 & 9.79$^{+0.15}_{-0.13}$ & $-2.5\pm0.2$ & $8.1^{+0.2}_{-0.3}$ & $1.3^{+1.0}_{-0.7}$ & $34^{+48}_{-25}$ & $-19.63$ & 03:32:42.13 & $-$27:46:50.28 & - & -  \\
22980 & 11.65$^{+0.17}_{-0.21}$ & $-2.3\pm0.5$ & $8.3^{+0.4}_{-0.6}$ & $2.4^{+2.5}_{-1.9}$ & $44^{+104}_{-36}$ & $-19.36$ & 03:32:39.55 & $-$27:46:28.62 & Yes & 11.58  \\
16748 & 11.77$^{+0.29}_{-0.44}$ & $-3.3\pm1.6$ & $7.9^{+0.4}_{-0.5}$ & $0.8^{+0.9}_{-0.6}$ & $41^{+115}_{-33}$ & $-18.71$ & 03:32:40.47 & $-$27:47:33.93 & Yes & -  \\
3216 & 12.56$^{+0.64}_{-0.66}$ & $-2.4\pm1.1$ & $8.7^{+0.4}_{-0.7}$ & $5.0^{+7.8}_{-3.9}$ & $40^{+107}_{-31}$ & $-19.13$ & 03:32:29.20 & $-$27:49:48.46 & - & -  \\
22450 & 13.54$^{+0.28}_{-0.76}$ & $-1.6\pm2.1$ & $7.8^{+0.4}_{-0.5}$ & $0.7^{+0.6}_{-0.5}$ & $52^{+87}_{-39}$ & $-18.73$ & 03:32:35.97 & $-$27:46:35.39 & - & 13.20  \\

		\hline
	\end{tabular}
\end{table*}
\citet{robertson2022} reported photometric properties of 4 galaxies with spectroscopic confirmation in a companion paper \citep{curtislake2022} using data from the JADES survey. We recover 3 of the 4 galaxies in their sample (the fourth source is positioned outside the medium band imaging) with our photometric redshifts broadly consistent with their spectroscopic redshifts. These are noted in the final column in Table \ref{tab:sample}. 

\citet{bouwens2022} also recently searched for galaxies using the JWST UDF medium band survey and found 4 galaxies at $z~\geq~9.5$. We recover 3 of these candidates in our sample as indicated in Table \ref{tab:sample}. The 4th source, XDFH-2334046578 in \citet{bouwens2022}, is not included in this sample as it failed the visual inspection stage due to an unusual positional offset of the source in F160W. Therefore we were not confident in the strength of the break between the F160W and F182M filters. For the 3 we recover, we find good agreement overall in the photometric redshifts. However, for UDF-22980 we measure a moderately lower redshift. We also find a significantly fainter absolute UV magnitude for this source of $M_{\rm{UV}}=-19.4$ compared to the value of $M_{\rm{UV}}=-20.2$ reported by \citet{bouwens2022}. Our value of $M_{\rm{UV}}$ is in close agreement to that given by \citet{robertson2022} which exploits the additional constraints provided by the broadband NIRCam imaging in the JADES survey. 

Our sample contains 2 galaxy candidates at $z>12.5$, UDF-3216 and UDF-22450 which have best fitting redshifts of $z=12.6^{+0.6}_{-0.7}$ and $z=13.5^{+0.3}_{-0.8}$ respectively. Despite the photometric redshift relying on detections in only 2 filters, our photometric redshift measurement of UDF-22450 is in close agreement with the spectroscopic redshift of $z_{\rm{spec}}=13.20$ recently reported by \citet{robertson2022}.

\subsection{Physical properties}
All the fluxes were corrected to total assuming a point source correction based on the curve-of-growth. We then use the \textsc{Bagpipes} spectral fitting code \citep{carnall2018} to further constrain the physical properties of the galaxies in our sample. We run \textsc{Bagpipes} using the configuration described in \cite{Carnall2020, carnall2022}, which includes the 2016 updated version of the \cite{bruzual2003} stellar population models with the MILES stellar spectral library, an emission line prescription calculated using the \texttt{Cloudy} photoionization code \citep{Ferland2017}, the \cite{Salim2018} dust attenuation model and a constant SFH model. The recovered stellar masses and mass-weighted mean stellar ages for all 6 galaxies are noted in Table \ref{tab:sample}. We find that these galaxies have relatively low stellar masses ($\log(M_{\star}/{\rm M_{\odot}}) \simeq 7.5 - 8.7$) and star-formation rates in the range  $\log(\rm{SFR}/M_{\odot}\rm{yr}^{-1}) \simeq 0.3 - 5.0$. The limited detections longward of the Balmer break prevent any conclusive constraints on the stellar ages of these galaxies. Nevertheless, the mean stellar ages derived from \textsc{Bagpipes} are moderate ($\sim 40$ Myr), and for those with spectroscopic confirmation, are consistent with ages derived using the more extensive filter coverage in the JADES survey \citep{robertson2022}.

Using the filters long-ward of the Lyman break we determine the rest-frame UV continuum slope, $\beta$ (where flux $f_{\lambda} \propto \lambda^{\beta}$) with the technique described in \citet{cullen2022}. Due to the limited number of filters beyond the Lyman break, the UV slope remains relatively unconstrained for most of these objects. However, we measure a robust UV slope for UDF-21003, finding $\beta=-2.5\pm0.2$. This represents a very blue UV continuum indicating the presence of a young, low-metallicity stellar population. However, we note that \citet{cullen2022} found a median UV slope of $\beta = -2.29 \pm 0.09$ at $8<z<16$, indicating that UDF-21003 is fully consistent with the average UV slope found at these redshifts.

\section{Luminosity function}
\label{sec:LF}
We compute the UV luminosity function at $z\simeq 11.2$ using objects detected at $8\sigma$ in F182M in a redshift bin spanning $9.5\leq z \leq 13.5$ (with the upper redshift limit set by the fact the Lyman break cuts into the F182M filter at $z\simeq13.5$). Four galaxies are selected using this condition which are contained within a bin centred at $M_{\rm{UV}}=-19.2$ with a width of 1 magnitude. The use of a redshift bin spanning $\Delta z = 4$ reduces the impact of photo-z uncertainties on our derivation of the LF.

The binned co-moving number density of sources per absolute magnitude, $\Phi(M_{\rm UV})$, is determined using the $1/V_{\rm max}$ method \citep{schmidt1968}. The volume, $V_{\rm{max}}$, is defined as the difference in co-moving volume between the volume at $z=9.5$ and the maximum redshift that the galaxy could be detected at $8\sigma$ in F182M with a limit at $z=13.5$ for this selection. We find that 2 of the sources are detectable to $z=13.5$ whereas the 2 fainter sources are only detectable to $z\simeq11.8$ and $z\simeq12.6$. Therefore the average redshift of the limited volumes is $z\simeq11.2$. In order to account for incompleteness in the galaxy sample, we perform completeness simulations using the same method as described in \citet{donnan2022}, with point sources injected into the F182M imaging in 3 regions across the 2 modules. From $m_{\rm{AB}}=24$-$31$ in steps of 0.1, 800 sources are injected into each region and the rate of recovery measured. This process is repeated 10 times for each region which therefore results in a total of 1,680,000 sources injected into the imaging. This generates the completeness as a function of apparent magnitude in the detection image, F182M. 

\begin{figure}
	\includegraphics[width=\columnwidth]{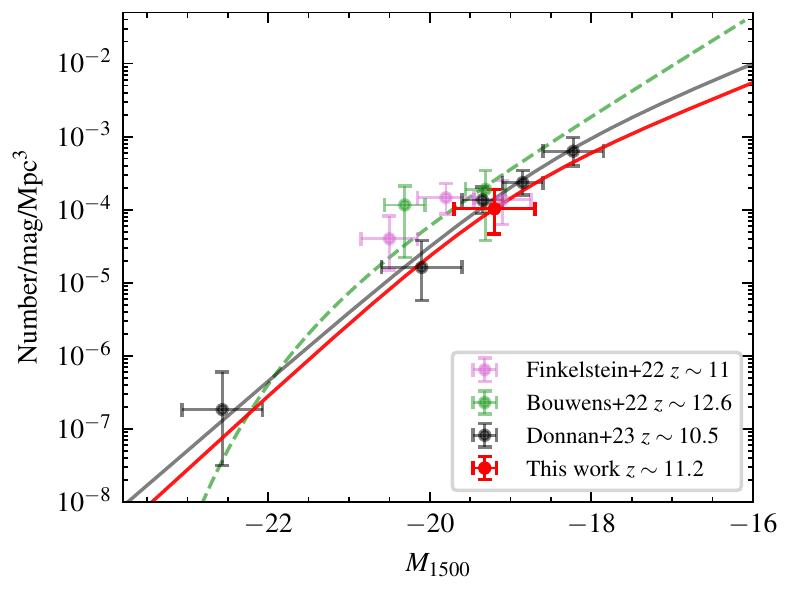}
    \caption{The UV luminosity function (LF) determined from our new sample in the redshift range $9.5<z<13.5$ (red point). The rest-frame UV LF at $9.5<z<11.5$ from \citet{donnan2022} is shown as black points  with the best-fitting double power-law shown as a solid black line. We include data points from \citet{finkelstein2022c} and \citet{bouwens2022}, with the best-fitting Schechter function from \citet{bouwens2022} plotted as the dashed green line. The derived number density based on our new $9.5<z<13.5$ galaxy sample is clearly consistent with the evolution of the UV LF derived by \citet{donnan2022}.} 
    \label{fig:LF}
\end{figure}

As the UV LF is derived here using a relatively small cosmological volume (\citet{donnan2022} use a $\sim 5 \times$ larger volume), our results are more vulnerable to cosmic variance. To account for this we use the cosmic variance calculator from \citet{trenti2008}. We measure an uncertainty of $27\%$ due to cosmic variance which is smaller than the uncertainties arrising from the small sample size. The cosmic variance uncertainty was factored into the final uncertainty on the number density. The resultant galaxy number density in this bin is shown in Fig. \ref{fig:LF} where $\Phi(M_{\rm UV}=-19.2) = (1.03^{+0.86}_{-0.57}) \times10^{-4}$ mag$^{-1}$ Mpc$^{-3}$.

\begin{figure*}
	\includegraphics[width=\textwidth]{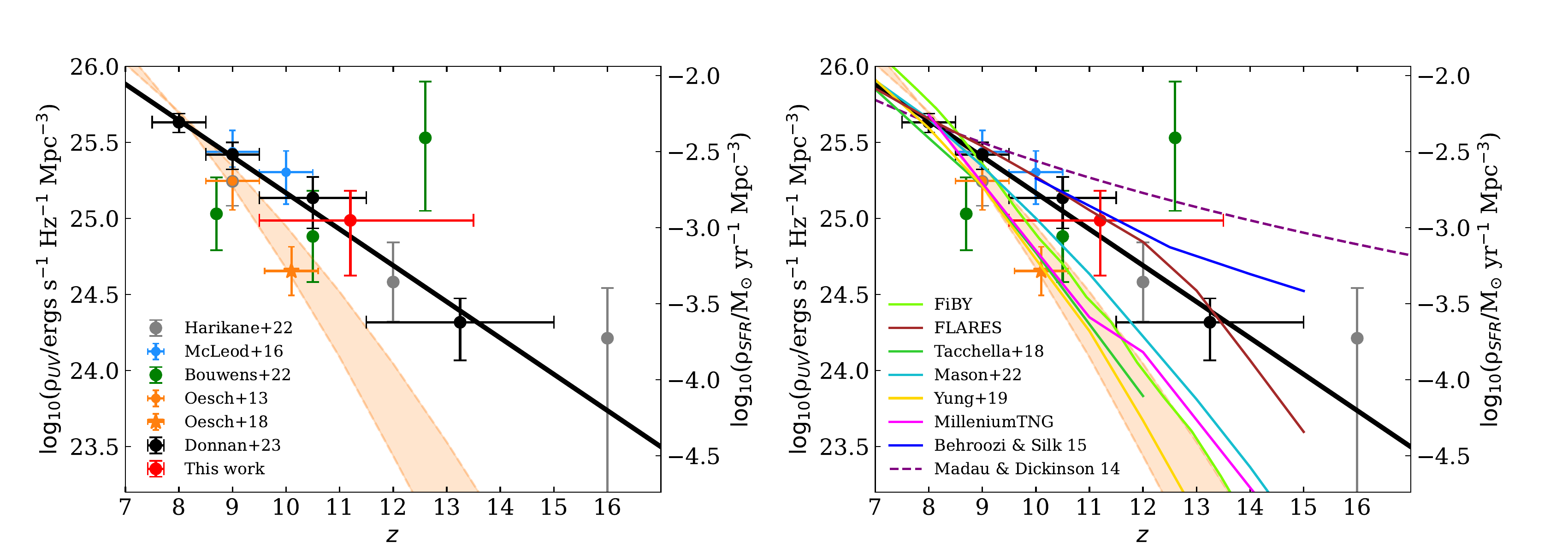}
    \caption{The redshift evolution of the UV luminosity density, $\rho_{\rm UV}$, and hence the inferred cosmic star-formation rate density, $\rho_{\rm SFR}$, at $z>7$ with our new measurement at $z\sim11.2$ (red circular data point). Estimates at $z\simeq9-10$ from \citet{oesch2013,oesch2018} and \citet{mcleod2016} are shown by the orange and blue data points respectively. The green and grey points show $\rho_{\rm UV}$ derived from the LFs in \citet{bouwens2022} and \citet{harikane2022} respectively. All values were determined using a limit of $M_{\rm UV}=-17$ in the luminosity-weighted integral. The black data points and solid line show the results from \citet[][]{donnan2022}. The shaded orange region shows the halo evolution model from \citet{oesch2018}. The right-hand panel shows a comparison to theoretical models, again setting the LF integration limit to $M_{\rm UV}=-17$. The green, cyan, blue and yellow lines show semi-analytic models from \citet{tacchella2018}, \citet{mason2022}, \citet{behroozi2015} and \citet{yung2019} respectively. The green, brown and pink lines show the results of hydrodynamical simulations from FiBY \citep{johnson2013,paardekooper2015}, FLARES \citep{wilkins2022} and MilleniumTNG \citep{kannan2022}. The dashed purple line shows an extrapolation at $z\geq8$ of the $\rho_{\rm{UV}} \propto (1+z)^{-2.9}$ relation from \citet{madau2014}. Our new measurement is fully consistent with the steady, exponential decline in $\rho_{\rm UV}$ up to $z\simeq15$ described in \citet{donnan2022}.} 
    \label{fig:rho_uv}
\end{figure*}

This new estimate at $z\simeq11.2$ is fully consistent with the recently-reported evolution of the UV LF from $z\sim10.5$ to $z\sim13.25$ \citep[][]{donnan2022}. We fit a double-power law (DPL) using the same technique described in \citet[][]{donnan2022} where we fix $\alpha$, $\beta$ and $M_{*}$ to $-2.10$, $-3.53$, $-19.12$ respectively. These are the best fitting values at $z\sim10.5$ \citep[][]{donnan2022} and therefore we allow only $\phi_{*}$ as a free parameter. This results in a best-fitting DPL model with $\phi_{*}=(2.36\pm1.60)\times10^{-4}$/mag/Mpc$^3$. We then perform a luminosity-weighted integral of the best-fitting DPL model to derive the comoving UV luminosity density ($\rho_{\rm{UV}}$) and therefore cosmic star-formation rate density ($\rho_{\rm{SFR}}$). This is shown as the red point in Fig. \ref{fig:rho_uv} which is in good agreement with the log-linear relation from \citet[][solid black line]{donnan2022} and is therefore consistent with a steady, exponential decline in $\rho_{\rm UV}$ to $z\sim15$. Our results are inconsistent with the result at $z\sim12.6$ from \citet{bouwens2022} as they predict a significantly higher $\rho_{\rm{UV}}$ as shown by the green point at $z\sim12.6$ in Fig. \ref{fig:rho_uv}. This is possibly impacted by their estimate of $M_{\rm{UV}}$ for UDF-22980 which is $\sim0.8$ mag brighter than derived here (see Section \ref{sec:sample}). They also restrict their search to the HUDF area alone and therefore probe a smaller cosmological volume than utilised in this search, which, as they mention, also makes their measurement of the UV LF somewhat more sensitive to cosmic variance. They report a cosmic variance uncertainty of $\sim52 \%$ using the \citet{trenti2008} calculator which is notably larger than we calculate for our LF ($\sim 27 \%$). \citet{bouwens2022} also use a significantly steeper value of the faint end slope of $\alpha = -2.71$. As $\rho_{\rm{UV}}$ is particularly sensitive to $\alpha$, their steeper faint-end slope also contributes to this increased value of $\rho_{\rm{UV}}$.

\section{Discussion}
\label{sec:discussion}
\subsection{The growth of early galaxies}
The early-release NIRCam data demonstrated the ability of \textit{JWST} to detect numerous galaxies at $z>9$ \citep[e.g.][]{adams2022b,castellano2022,naidu2022,finkelstein2022} including a detection of a galaxy candidate at $z\simeq16$ \citep{donnan2022} which has since been verified by multiple studies \citep[e.g.][]{harikane2022,finkelstein2022c}. This indicates that galaxy formation commenced at $z>16$ and therefore within < 250 Myr of the big bang. The results from this new study give further support to this conclusion with the detection of 6 galaxies at $9.5<z\lesssim13.5$ from within a relatively small cosmological volume. As demonstrated in this analysis, this corresponds to a number density consistent with the evolution of the UV LF derived by  \citet{donnan2022}. This leads to a value of the UV luminosity density (and cosmic star-formation rate density) at $z\simeq11.5$ which is fully consistent with the log-linear relation derived in \citet{donnan2022} of:

\begin{equation}
\rm{log}_{10}(\rho_{\rm UV}) = (-0.231\pm0.037) z + (27.5\pm 0.3).
\end{equation}

The results of this study thus further support the analytical prediction from \citet{hernquist2003} that a steady, exponential decline in $\rho_{\rm{SFR}}$ is expected with increasing redshift, as illustrated by the solid line in the left-hand panel of Fig. \ref{fig:rho_uv}. In the right-hand panel of Fig. \ref{fig:rho_uv} we additionally provide a comparison of the new observational constraints on the UV luminosity density with the  predictions of various different theoretical/numerical models. Although the observational data points lie below the extreme-redshift extrapolation of the relation provided by  \citet{madau2014}, they sit above the predictions from semi-empirical constant star-formation efficiency models as deduced by \citet{mason2022} and \citet{tacchella2018}. We also compare to the cosmological hydrodynamical simulations FiBY \citep{johnson2013,paardekooper2015}, FLARES \citep{wilkins2022} and MilleniumTNG \citep{kannan2022}. Overall, it can be seen that there is a wide range in the predictions of the different theoretical models; the new observational constraints sit above most of the theoretical curves but are reasonably in-line with the predictions from FLARES and the semi-analytical model presented by \citet{behroozi2015}.

\subsection{Low-redshift solutions for the $\mathbf{z\simeq12}$ candidate}
After UDF-22980 was proposed to lie at $z=11.9$ \citep[][]{ellis2013,mclure2013}, there followed much discussion on the nature of this object, in part informed by a series of follow-up observations. \citet{brammer2013} used \textit{HST} GRISM spectroscopy and \citet{capak2013} used Keck MOSFIRE to search for emission lines. No robust emission lines were detected but tentative ($\sim2 \sigma$) detections were reported at $\lambda_{\rm{obs}} \simeq 1.6 \mu$m. It was concluded that this could be consistent with an [OIII]$\lambda$4959,5007 detection at $z=2.19$ or  [OII]$\lambda$3727,3729 at $z=3.29$. Now, however, with the \textit{JWST} medium band imaging at $\lambda_{\rm{obs}}>1.6\,\mu$m we can robustly constrain the shape of the galaxy SED. We find that, fitting with \texttt{EAZY} and \texttt{LePhare}, the new enhanced photometry is inconsistent with any solution at at $z=2.19$ or $z=3.29$. When the fit is forced to these redshifts we find that the best-fitting (formally unacceptable) solution is an old passive galaxy which would not have the strong emission lines required for these low-z solutions. Another inconsistency with the low-z solution is that a strong [OIII] emitter at $z=2.19$ would likely have strong H$\alpha$ emission at $\lambda_{\rm{obs}}=2.1\,\mu$m. This would result in a substantially smaller flux in F182M than in F210M which is not what is observed. This conclusion is also reinforced by the recent {\it JWST} spectroscopic observations from the JADES survey, albeit no emission lines were detected with NIRSpec \citep{robertson2022,curtislake2022}.

\subsection{The power of medium-band imaging}
Accurate spectroscopic confirmation of $z$ $\geq$ $7$ galaxies at optical/near-infrared wavelengths has been, and indeed continues to be a challenge due to the attenuation of Ly$\alpha$ emission within the epoch of reionization. An alternative route to robust spectroscopic redshifts is provided by the Atacama Large Millimeter/submillimeter Array (ALMA), with spectroscopic redshifts of $z\geq7$ galaxies now regularly secured at sub-mm wavelengths. For example, ALMA has now been used to confirm galaxies at $z=9.1$ \citep{hashimoto2018} and $z=8.38$ \citep{laporte2017} based on the detection of [OIII] at $\lambda_{\rm{rest}} = 88\,\mu$m as well as observations of [CII] at $\lambda_{\rm{rest}} = 158\,\mu$m. As the required observation time with ALMA is strongly dependent on the confidence interval of the photometric redshift (due to the potential requirement for multiple tunings), robust photometric redshifts are required for efficient sub-mm/mm spectroscopic confirmation. The results presented here thus demonstrate that \textit{JWST} NIRCam medium-band imaging can be an invaluable additional  tool for refining the photometric redshifts of high-redshift galaxy candidates in preparation for ALMA follow-up spectroscopy.

\section{Conclusions}
\label{sec:conclusions}
We have used a combination of the deepest optical + near-infrared \textit{HST} broad-band imaging with deep \textit{JWST} NIRCam medium-band imaging to detect 6 galaxy candidates at $z\geq9.5$. 

We obtain a robust photometric redshift for 6 candidates including the previously reported $z\simeq12$ galaxy candidate, UDF-22980, at $z_{\rm{phot}}\simeq 11.6$. We find that the previously suggested low-z emission line galaxy solutions indicated from $\sim2 \sigma$ spectroscopic detections at $\lambda_{\rm{obs}}=1.6\, \mu$m are incompatible with the \textit{JWST} NIRCam photometry which instead re-affirms the $z\simeq12$ solution and refines the photometric redshift to $z = 11.65^{+0.17}_{-0.21}$. The robustness of the solutions were tested using 4 independent SED fitting codes which all show good agreement. 

We also report UV slopes, stellar masses, star-formation rates and mass-weighted mean stellar ages for all 6 candidates. We find stellar masses of $\log(M_{\star}/{\rm M_{\odot}}) \simeq 7.5 - 8.7$ and star-formation rates of $\log(\rm{SFR}/M_{\odot}\rm{yr}^{-1}) \simeq 0.3 - 5.0$. We are able to measure a robust UV continuum slope for one object, finding a very blue slope of $\beta=-2.5\pm0.2$ which is consistent with a young, low-metallicity stellar population.

The galaxy number density and hence cosmic star-formation rate density at $z\simeq11$ derived from this sample are fully consistent with the current picture of the evolution of the UV LF with a gradual steady decline over the redshift range $z=8$-$15$. The resulting inferred gradual (exponential) decline in cosmic star-formation rate density out to at least $z \simeq 15$ is clearly inconsistent with at least some theoretical/numerical models of early galaxy evolution.

This study further demonstrates the power of \textit{JWST} to quickly reveal the true nature of $z\geq10$ candidates that were first identified by \textit{HST} as well as to detect new galaxy candidates, with the NIRCam medium-band imaging analysed here providing sensitive photometry at $\lambda_{\rm{obs}}\gtrsim 1.6\, \mu$m. Robust photometric redshifts are vital for drawing conclusions about the physical properties of $z\geq10$ galaxies as well as providing a tighter frequency range for ALMA follow-up spectroscopy. Forthcoming \textit{JWST} surveys will primarily target fields that are accessible to ALMA, and were heavily imaged by \textit{HST} such as NGDEEP (GO 2079) and PRIMER (GO 1837)\footnote{https://primer-jwst.github.io/} and will therefore benefit from this combination. Therefore, it is likely these surveys will provide large statistically-significant samples of galaxies over a wide redshift range which will yield further constraints on the high-redshift evolution of the UV luminosity function as well as providing numerous extreme-redshift galaxy candidates for follow-up {\it JWST} and ALMA spectroscopy.

\section*{Acknowledgements}

We thank the referee Richard S. Ellis for useful comments which helped improve the quality of the manuscript. We are grateful to Stephen Wilkins and Andrea Incatasciato for providing simulation data. C. T. Donnan, D. J. McLeod, R. J. McLure, J. S. Dunlop acknowledge the support of the Science and Technology Facilities Council. 
A.C. Carnall thanks the Leverhulme Trust for their support via the Leverhulme Early Career Fellowship scheme.
F. Cullen acknowledges support from a UKRI Frontier Research Guarantee Grant [grant reference EP/X021025/1]. For the purpose of open access, the author has applied a Creative Commons Attribution (CC BY) licence to any Author Accepted Manuscript version arising from this submission.

%%%%%%%%%%%%%%%%%%%%%%%%%%%%%%%%%%%%%%%%%%%%%%%%%%
\section*{Data Availability}

The inclusion of a Data Availability Statement is a requirement for articles published in MNRAS. Data Availability Statements provide a standardised format for readers to understand the availability of data underlying the research results described in the article. The statement may refer to original data generated in the course of the study or to third-party data analysed in the article. The statement should describe and provide means of access, where possible, by linking to the data or providing the required accession numbers for the relevant databases or DOIs.



\bibliographystyle{mnras}
\bibliography{udf_paper} % if your bibtex file is called example.bib

%\begin{thebibliography}{99}
%\end{thebibliography}

%%%%%%%%%%%%%%%%%%%%%%%%%%%%%%%%%%%%%%%%%%%%%%%%%%

%%%%%%%%%%%%%%%%% APPENDICES %%%%%%%%%%%%%%%%%%%%%

%\appendix

%%%%%%%%%%%%%%%%%%%%%%%%%%%%%%%%%%%%%%%%%%%%%%%%%%

% Don't change these lines
\bsp	% typesetting comment
\label{lastpage}
\end{document}